\documentclass[seceq,supplement]{ptptex}

\usepackage{graphicx}

\title{
Long-time tail in 
an 
electric conduction system}

\author{
Tatsuro \textsc{Yuge} and Akira \textsc{Shimizu}
}

\inst{
Department of Basic Science, 
University of Tokyo, 3-8-1 Komaba, Meguro-ku, Tokyo 153-8902, Japan
}

\abst{
The long-time behavior of the velocity autocorrelation function 
in a classical two-dimensional electric conduction system is studied 
by the molecular dynamics simulation.
In equilibrium, 
the effect of coexistence of many-body interactions 
and a random potential is investigated.
A crossover from a positive tail proportional to $t^{-1}$, 
to a negative tail proportional to $-t^{-2}$ is observed
as the strength of the random potential increases.
In nonequilibrium, 
the positive tail is enhanced 
whereas the negative tail appears at earlier times 
as an electric field increases.
}

\begin{document}

\maketitle

\section{Introduction}

The long-time tail was 
first discovered 
by B. J. Alder and T. E. Wainwright\cite{AW} 
for the velocity autocorrelation function (VACF) 
in a hard-core fluid system.
This tail is positive and proportional to $t^{-d/2}$
(hereafter called the fluid-type tail)
\cite{EHL1,DC1,ZB,Kawasaki1,PomeauResibois,KBS}.
Here, $d$ is the dimension of the system.
As a result the self-diffusion coefficient $D$ in this system 
is logarithmically divergent in terms of the system size $L$ for $d=2$.

Another system which has a long-time tail 
is the Lorentz model, 
which describes the motion of a single particle in a disordered system.
In this system the tail of the VACF is negative 
and proportional to $-t^{-(d+2)/2}$
(called the Lorentz-type tail hereafter)\cite{EW,Bruin1,AA1,AA2,GLY1,EMDB1}.
Although $D$ is not divergent even for $d=2$,
this tail might cause an algebraically $L$-dependent term in $D$.

In the hard-core fluid only a many-body interaction exists 
and in the Lorentz model only a random potential exists.
Hence 
a question occurs:
{\it How does the long-time tail change 
when a many-body interaction and a random potential coexist?}
A typical system which has both a many-body interaction 
and a random potential 
is an electric conduction system.
In this system electron-electron ({\it e-e}) interaction corresponds to 
a many-body interaction 
and electron-impurity ({\it e-i}) interaction to a random potential.

In studies of the long-time tails in the hard-core fluid and the Lorentz model
the equilibrium states have mainly been 
investigated.
Regarding electric conduction systems, 
another interesting question arises:
{\it How does the long-time tail change under a nonequilibrium condition?}

In this paper we report the results of 
the molecular dynamics (MD) simulations 
on a model of 
a two-dimensional 
electric conduction system 
to answer these two questions\cite{YS}.

\section{Model and its physical meanings}

We use a model of electric conduction proposed in Ref.~\citen{YIS}.
We here explain its physical meanings in more detail.

The model is a two-dimensional classical system, 
the size of which is $L_x \times L_y$.
In the system are three sorts of particles, 
which we call electrons, phonons and impurities\cite{note0}.
An external electric field $E$ is applied in the $x$-direction, 
which acts only on electrons.
The boundary condition in the $x$-direction is periodic, 
and that in the $y$-direction is a potential wall for electrons 
and a thermal wall with temperature $T$ for phonons.
Moreover we assume that short-range interactions are present
among {\em all} these particles.
The charge of an electron is denoted by $e$\cite{note0}.
The mass, radius, number density of the electrons (phonons)
are denoted by $m_e$ ($m_p$), $R_e$ ($R_p$) and $n_e$ ($n_p$), respectively.
The impurities are immobile and play the role of a random potential.
The radius and number density of the impurities 
are denoted by $R_i$ and $n_i$, respectively.
The configuration of the impurities is given by an almost uniform distribution 
except for the restriction that the distance between any pair of 
the impurities is larger than $2(R_e+R_i)$.
The initial positions of the electrons and phonos are 
randomly arranged 
not to overlap with the other particles, 
and their initial velocities are given by the Maxwell distribution 
with temperature $T$.

We can control the frequencies of the {\it e-e} 
and {\it e-i} collisions by changing $n_e$ and $n_i$.
Thus we can investigate independently the effects 
of the many-body 
interactions ({\it e-e}, {\it e-p} and {\it p-p} interactions) 
and that of the random potential ({\it e-i} interaction)
on the long-time behavior of the VACF.
It should be also noted 
that the situation where $n_i=0$ 
corresponds to the hard-disk fluid 
and that the situation without {\it e-e} interaction
corresponds to the (non-overlapping) Lorentz model.

A typical system corresponding to this model is 
a two-dimensional electron system in 
a doped semiconductor at room 
temperature.
Since the Fermi energy is smaller than room temperature
in such a system, 
electrons can be treated as classical particles.
Furthermore, 
we can treat them as being confined in a two-dimensional plane 
because the temperature is lower than the 
exciting energy 
between the ground and the second subbands\cite{subband} of an electron.

It is well-known
that in a {\em uniform} solid (such as our model), 
the motion of the electron system 
can be separated into
the collective oscillations and the individual motions, 
and that the long-range effects of the Coulomb force 
are adequately incorporated in the collective oscillations\cite{Pines}. 
Moreover the collective oscillations do not play the central 
role in the electric conduction.
[It is worth mentioning here that 
for {\em non-uniform} systems, such as a conductor connected to 
electron reservoirs\cite{SM1998,SK2000}, 
the long-range effects 
should be treated more carefully as discussed 
in Refs.~\citen{SM1998,SK2000}.]
For these reasons, we can treat the electrons 
as individual particles interacting through a short-range force
when discussing transport properties of conductors.
We can estimate the range (the screening length) of 
the effective interaction among electrons 
by the Debye length, $(k_{\rm B} T /4\pi n_e e^2)^{1/2}$, 
in such a classical system.
The potential produced by an impurity is also screened 
and its effective potential range is also estimated by the Debye length.
Therefore the interaction ranges of {\it e-e} and {\it e-i} interactions 
are comparable.

In real solids 
phonons are the oscillation modes of the crystal lattice in the conductors. 
Therefore the total number of phonons does not conserve.
However, because almost all the possible modes of phonos are excited 
in semiconductor 
at room 
temperature, 
the number density of phonons is so high 
that the non-conservation of the phonon number would be irrelevant. 
Moreover, the energy-momentum dispersion relations of phonons 
are complicated in real solids.
This would be also irrelevant, however, 
when discussing general nonequilibrium properties, 
which are independent of the details of the materials, 
of electric conduction.
In this study, therefore, 
it is sufficient to model phonons as classical particles\cite{particles} 
whose number conserves and 
whose mass is constant, which corresponds to parabolic dispersion relation.
Note that 
if one wants to reproduce the $T$-dependence 
of conductivity, such as the famous linear-$T$ dependence,
then the number of phonons should be 
varied as a function of $T$.
In this paper, however, we are not interested in the $T$-dependence 
of conductivity.

This model contains what we believe to be essential elements
of electric conduction in the context of nonequilibrium statistical physics.
These elements are the following: 
(i) A driving force which induces electric current: {\it electric field}. 
(ii) Careers which transfer heat to outside the conductor: 
{\it phonons}.
In real physical systems (i.e., in experiments), a 
conductor is surrounded by a large insulating material,
which works as the heat bath.
The energy supplied from an external electric field to electrons 
is dissipated 
as the Joule heat transferring into the heat bath 
through the walls of the conductor 
to keep the system in a steady state.
The heat flow {\em across} the walls of the conductor 
is mediated not by electrons but by phonons, 
while the heat flow {\em in} the conductor is mediated by both of them.
Therefore the electron-phonon interactions 
and heat contact of phonons at the walls are essential
to realize a nonequilibrium steady state.
(iii) Objects which break the microscopic 
translational invariance in the bulk region: 
{\it impurities}.
Although the conductor is macroscopically uniform, 
the microscopic translation invariance is violated 
by the imperfections of the conductor (impurities, defects and so on).
This defines the rest frame of the electrons at equilibrium,
and thereby eliminates any possible anomalies which may arise from 
the translational invariance.
(iv) Many-body interactions: {\it short-range interactions among all particles}.
(v) A nonequilibrium steady state is uniquely determined
by a small number of macroscopic parameters: 
$T$ {\it and} $E$.

We perform the time-step-driven MD on this model.
We set $R_e$, $m_e$, $k_{\rm B}T$ and $e$ to unity, 
and take $L_x=2000$, $L_y=125$, $R_p=1$, $R_i=0.5$ and $m_p=1$
in these units\cite{note:R}.
The number densities of the electrons and phonons are fixed to 
$n_e = 0.04$ and $n_p = 0.002$.
Since $n_e$ is taken larger than that in our previous study\cite{YS},
the long-time tails are more clearly observed.
In this study we employ the Hertzian interaction 
$U_{jk}^{\rm int}= Y(\max\{ 0, d_{jk}\})^{5/2}$
as the interaction potential between any pair of the particles.
Here, $Y$ is a constant (fixed to 4000 in the simulation), and 
$d_{jk} = R_j + R_k - |\mbox{\boldmath$r$}_j-\mbox{\boldmath$r$}_k|$ 
is the overlap of the potential ranges of the $j$-th and $k$-th particles
($R_j$ is the radius of the $j$-th particle and equals $R_e$, $R_p$ or $R_i$. 
$\mbox{\boldmath$r$}_j$ is the position of the $j$-th particle)\cite{note:R}.

\begin{figure}[t]
\begin{center}
\includegraphics[width=\linewidth]{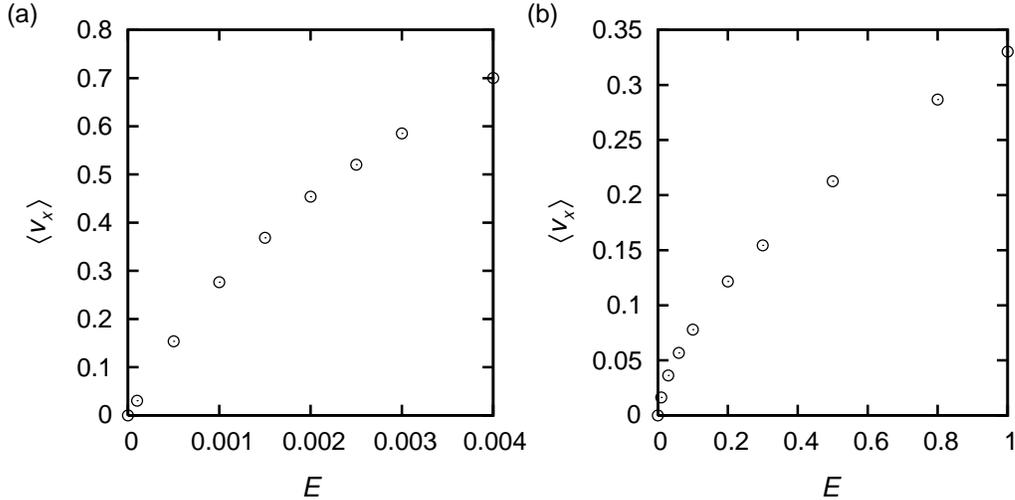}
\caption{
Average velocities of an electron for (a)~$n_i=0.0004$ 
and (b)~$n_i=0.04$, plotted against $E$.
}
\label{averageVelocity}
\end{center}
\end{figure}

We previously confirmed that 
this model shows good properties of steady states\cite{YIS}.
For example, as shown in Fig.~\ref{averageVelocity}, 
we observe
linear response of $v_x$ to $E$ near equilibrium 
and nonlinear response far from equilibrium,
where $v_x$ is the component of  
electron's velocity 
in the $x$-direction (parallel to $E$).
We also observed that 
the Kramers-Kronig relation of the complex admittance 
as well as 
the fluctuation-dissipation theorem\cite{Nyquist} 
hold at all frequencies in the equilibrium state.

\section{Results}

We calculate the VACF, 
\begin{equation}
C(t) = 
{\bigl\langle (v_x(t)-\langle v_x \rangle) 
(v_x(0)-\langle v_x \rangle) \bigr\rangle
\over \bigl\langle (v_x-\langle v_x \rangle)^2 \bigr\rangle},
\end{equation}
of an electron in steady states under various conditions.

\subsection{Equilibrium case}

\begin{figure}[t]
\begin{center}
\includegraphics[width=0.75\linewidth]{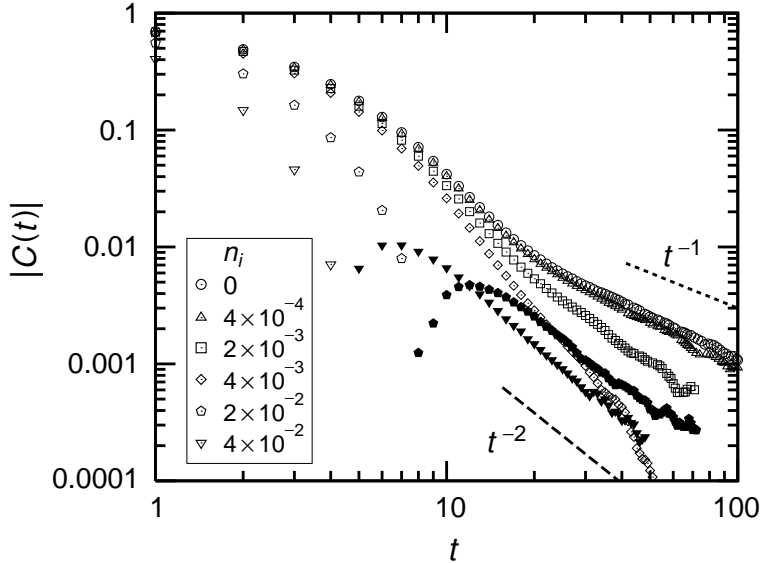}
\caption{
A double logarithmic plot of the absolute value of the VACF 
in the equilibrium states for various values of $n_i$.
The open and closed symbols correspond 
to positive and negative values of the VACF, respectively.
The dotted and dashed lines are reference lines proportional 
to $t^{-1}$ and $t^{-2}$, respectively.
}
\label{e10000eq}
\end{center}
\end{figure}

First we show the results of the VACF in the equilibrium states.
We calculate the VACFs with $n_e$ and $n_p$ fixed, 
and $n_i$ varied from 0 to 0.04.
In this case, we can investigate the effect of the random potential
introduced into the fluid system.
Figure \ref{e10000eq} displays the results. 
We observe power-law tails of the VACFs at longer times.
When $n_i=0$ the tail is positive 
and the exponent is approximately $-1$\cite{IsobeLTT}, 
which is the result for the fluid system.
As $n_i$ increases, the fluid-type tail becomes weaker, 
and eventually disappears.
Instead, we observe another tail which is negative 
and whose exponent is approximately $-2$.
That is, the long-time behavior of the VACF is a crossover 
from the fluid-type tail to the Lorentz-type tail
as $n_i$ becomes larger.

According to phenomenological 
theories\cite{EHL1,ZB,Kawasaki1,PomeauResibois,KBS,GLY1,EMDB1}, 
the origin of the long-time tail in a fluid is the mode coupling of 
the momentum field and the density field 
(because of the conservation of the total momentum and total number 
of the particles) 
while that in the Lorentz model is the coupling of the density field 
and a static mode generated by the configuration of the impurities.
From these theories, 
we have a qualitative explanation of the crossover from 
the fluid-type tail to the Lorentz-type tail.
When $n_i$ is small, 
the momentum conservation is approximately valid 
and we can regard the system as a fluid.
Then the momentum field can contribute to the hydrodynamic mode 
and we observe the fluid-type tail.
As $n_i$ increases, 
the violation of the momentum conservation becomes larger 
and simultaneously the static mode by the impurity configuration 
becomes more relevant.
Then the fluid-type tail disappears 
and the Lorentz-type tail appears.

A necessary condition to treat the electron system as a fluid 
is that the mean number $\bar{N}_e$ of the electrons 
in an area among the impurities is sufficiently large.
Because the mean area per impurity is $1/n_i$, 
$\bar{N}_e$ is estimated as $\bar{N}_e = n_e/n_i$.
Therefore we can use the ratio of the electron and impurity density 
as a criterion for the long-time behavior of the VACF.
From Fig.~\ref{e10000eq} the crossover value $(n_e/n_i)_c$ is evaluated as 
$2 \lesssim (n_e/n_i)_c \lesssim 10$ 
[the fluid-type tail appears when $n_e/n_i > (n_e/n_i)_c$ 
and the Lorentz-type tail appears when $n_e/n_i < (n_e/n_i)_c$].

\subsection{Nonequilibrium case}

\begin{figure}[t]
\begin{center}
\includegraphics[width=\linewidth]{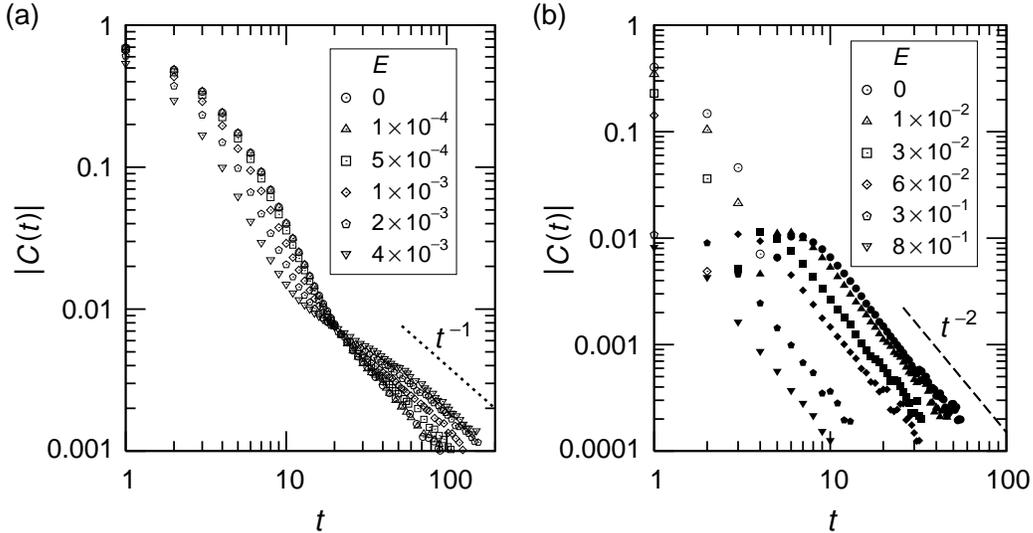}
\caption{
A double logarithmic plot of the absolute value of the VACF 
in the steady states for various values of $E$.
The open and closed symbols correspond 
to positive and negative values of the VACF, respectively.
The dotted and dashed lines are reference lines proportional 
to $t^{-1}$ and $t^{-2}$, respectively.
(a)~$n_i=0.0004$ [$n_e/n_i \gg (n_e/n_i)_c$; 
a corresponding $E$-$\langle v_x\rangle$ plot is 
in Fig.~\ref{averageVelocity}(a)].
(b)~$n_i=0.04$ [$n_e/n_i \ll (n_e/n_i)_c$; 
a corresponding $E$-$\langle v_x\rangle$ plot is 
in Fig.~\ref{averageVelocity}(b)].
}
\label{e10000noneq}
\end{center}
\end{figure}

Next, we show the results of the VACF when electric current is flowing.
We calculate the VACFs for $n_e/n_i \gg (n_e/n_i)_c$ and 
$n_e/n_i \ll (n_e/n_i)_c$, with varying $E$.
In Fig.~\ref{e10000noneq}, 
we show the long-time behavior of the VACFs of the steady states 
including the states in the nonlinear response regimes 
as well as those in equilibrium and in the linear response regimes.
When $n_i=0.0004$ [$n_e/n_i \gg (n_e/n_i)_c$], we observe 
that the amplitude of the fluid-type tail gets enhanced 
as $E$ increases.
Because $C(t)$ is normalized 
by $\bigl\langle (v_x-\langle v_x \rangle)^2 \bigr\rangle$ 
to be $C(0)=1$, 
this enhancement is not simply due to 
the rise of the kinetic temperature of the electrons.
This result implies that 
a ``temperature'' is different among different time scales 
in nonequilibrium states.
For the system with $n_i=0.04$ [$n_e/n_i \ll (n_e/n_i)_c$], 
the Lorentz-type tail appears at earlier times 
as $E$ becomes larger.
One of the reasons for the early appearance of the Lorentz-type tail 
would be that the duration time of the {\it e-i} collision 
becomes shorter as $E$ increases.

\section{Concluding Remarks}

In summary, we have investigated the long-time behavior of the VACF
in a system where many-body interactions and a random potential coexist.
In equilibrium, we have observed 
a crossover from the fluid-type tail to the Lorentz-type tail
as the impurity density increases.
We have interpreted that this crossover occurs 
because the system can not be regarded as a fluid 
when the impurity density becomes large.
The ratio of the electron and impurity densities 
is a criterion quantity for this crossover.
In nonequilibrium, 
as an electric field increases we have observed that 
the fluid-type tail is enhanced for $n_e/n_i \gg (n_e/n_i)_c$
and that the Lorentz-type tail appears 
at earlier times for $n_e/n_i \ll (n_e/n_i)_c$.

Finally, we list 
some issues related to this study.

(1) The long-time tails in equilibrium states
induce system-size dependence of $D$.
In an electric conduction system $D$ can be translated 
into the electrical conductivity $\sigma$ in the linear response regime
by the Einstein relation.
Thus some system-size dependence of $\sigma$ 
should be observed in experiments.
To our knowledge, however, 
no experiment has been reported 
which observed 
a system-size dependence of $\sigma$
for uniform two-dimensional electron systems 
at room temperature\cite{note:localization}.
This might be because the amplitude of the system-size dependent term 
in $\sigma$ is too small to detect 
in comparison with the system-size independent term,
or because phonon scattering would introduce 
a system-size independent cutoff.

(2) In Sec.~3.1 we have presented a possible scenario 
of a mode-coupling theory for the crossover 
from the fluid-type tail to the Lorentz-type tail
in equilibrium.
This should be explicitly shown 
by combining the mode-coupling theories 
for the fluid\cite{EHL1,ZB,Kawasaki1,PomeauResibois,KBS} 
and for the Lorentz model\cite{GLY1,EMDB1}.

(3) This crossover behavior would also be 
supported by a kinetic theory 
with a two-parameter ($n_e$ and $n_i$) expansion, 
which combines the kinetic theory 
for the hard-core fluid ($n_e$-expansion)\cite{DC1}
and 
that
for the Lorentz gas  ($n_i$-expansion)\cite{EW}.

(4) A crossover might occur 
from the fluid-type tail to the Lorentz-type tail, 
even in a system with lower density of impurities 
[that is, $n_e/n_i > (n_e/n_i)_c$], 
at longer times when the effect of the violation 
of the momentum conservation becomes relevant.
If this is true, $D$ is convergent\cite{note:Burnett} 
in two-dimensional systems 
except when the impurity density vanishes.
To demonstrate this, a larger-scale simulation 
with higher accuracy is necessary.
This issue should also be studied by a mode-coupling theory 
and a kinetic theory.

(5) In Sec.~3.2 we have observed an enhancement of the fluid-type tail 
in a nonequilibrium steady state.
This might be explained by a mode-coupling theory 
similar to 
the
one recently developed in sheared fluids\cite{OH}.

\section*{Acknowledgements}
We would like to thank N. Ito for helpful discussions
and also thank many of the participants in the symposium 
for giving useful comments.
This work was supported by a Grant from the Research
Fellowships of the Japan Society for the Promotion of
Science for Young Scientists (No. 1811579),
and by KAKENHI (No. 19540415).


\begin{thebibliography}{99}



\bibitem{AW}
B. J. Alder and T. E. Wainwright, 
Phys. Rev. A  {\bf 1} (1970), 18.


\bibitem{EHL1}
M.~H. Ernst, E.~H. Hauge and J.~M.~J. van Leeuwen, 
Phys. Rev. Lett. {\bf 25} (1970), 1254.

\bibitem{DC1}
J.~R. Dorfman and E.~D.~G. Cohen, 
Phys. Rev. Lett. {\bf 25} (1970), 1257.

\bibitem{ZB}
R. Zwanzig and M. Bixon, 
Phys. Rev. A {\bf 2} (1970), 2005.

\bibitem{Kawasaki1}
K. Kawasaki, Phys. Lett. A {\bf 32} (1970), 379.

\bibitem{PomeauResibois}
Y. Pomeau and P. R{\'e}sibois, 
Phys. Rep. {\bf 19} (1975), 63.

\bibitem{KBS}
 T.~R. Kirkpatrick, D. Belitz and J.~V. Sengers, 
J. Stat. Phys. {\bf 109} (2002), 373.


\bibitem{EW}
M.~H. Ernst and A. Weijland, 
Phys. Lett. A {\bf 34} (1971), 39.

\bibitem{Bruin1}
C. Bruin, Phys. Rev. Lett. {\bf 29} (1972), 1670.

\bibitem{AA1}
B.~J. Alder and W.~E. Alley, 
J. Stat. Phys.{\bf 19} (1978), 341.

\bibitem{AA2}
B.~J. Alder and W.~E. Alley, 
Physica {\bf 121A} (1983), 523.

\bibitem{GLY1}
W. G{\"o}tze, E. Leutheusser and S. Yip, 
Phys. Rev. A {\bf 23} (1981), 2634.

\bibitem{EMDB1}
M.~H. Ernst, J. Machta, J.~R. Dorfman and H. van Beijeren, 
J. Stat. Phys. {\bf 34} (1983), 477.


\bibitem{YS}
T. Yuge and A. Shimizu, J. Phys. Soc. Jpn. {\bf 76} (2007), 093001;
erratum {\bf 77} (2008), 028001.


\bibitem{YIS}
T. Yuge, N. Ito and A. Shimizu,
J. Phys. Soc. Jpn. {\bf 74} (2005), 1895.


\bibitem{note0}
We set the charge of the electrons positive.
As usual in the solid state physics, 
the back ground charge (negative in this case) 
is included in the system, 
which ensures the charge neutrality of the system.


\bibitem{subband}
In a real two-dimensional electron system,
electron motion in the confined ($z$) direction is quantized.
In {\em each} quantized level $n =1, 2, \cdots$, 
a conduction electron moves freely 
in the {\it x-y} plane, and thus a small band is formed 
which is called a subband.
The conduction band therefore splits into subbands 
which are labeled by $n$.
Conduction electrons can be well regarded as two-dimensional 
electrons when only the ground subband is occupied.


\bibitem{Pines}
D.~Pines, Solid State Physics {\bf 1} (1955), 367.


\bibitem{SM1998}
A. Shimizu and T. Miyadera, Physica B {\bf 249}-{\bf 251} (1998), 518.

\bibitem{SK2000}
A. Shimizu and H. Kato, 
Nonequilibrium Mesoscopic Conductors Driven by Reservoirs,  
Low-Dimensional Systems --- interactions and Transport Properties 
[ed. T. Brandes, Springer, 2000], pp.3-22 (arXiv:cond-mat/9911333).




\bibitem{particles}
One might suspicious about treating waves as particles.
However, it can be shown that a classical mixture (mixed state), 
such as the one treated here,
of many extended quantum modes (plane waves)
is identical to a classical mixture of
localized quantum wavepackets.
A wavepacket can be treated as a classical particle
in a classical regime.
Furthermore, in a doped semiconductor at room temperature
the average wavelength of phonons is shorter than 
that of conduction electrons because 
width of the phonon band
$\sim k_B T \ll$
 width of the conduction band.
Hence, the size of a wavepacket of a phonon is 
shorter than that of an electron.
It is well-known that at room temperature an electron can be treated as 
a classical particle (with respect to the motion in 
the {\it x-y} (non-confining) plane for the case of two-dimensional
electron systems).
Therefore, a phonon can also be treated as 
a classical particle.
The difference of the statistics (fermion versus boson) 
is not important at such a high temperature.


\bibitem{Nyquist}
H. Nyquist, Phys. Rev. {\bf 32} (1928), 110.


\bibitem{note:R}
Note that $R_e$, $R_p$ and $R_i$ are not the real radii 
but the effective lengths of the interaction ranges.


\bibitem{IsobeLTT}
When the packing fraction of the fluid particles is high, 
the exponent slightly deviates from $-1$ for hard-disk fluid.
See, M. Isobe, Phys. Rev. E {\bf 77} (2008), 021201.


\bibitem{note:localization}
Note that although the weak (Anderson) localization effect 
causes a system-size dependence of $\sigma$ at low temperature,
the origin of this dependence is different from 
the long-time tails in classical systems 
and this effect becomes weaker at room temperature.
It is also an interesting problem how the effect of the long-time tail 
in a classical system and the weak localization effect in a quantum system
on $\sigma$ are connected.
A similarity between the fluid-type tail and the weak localization effect 
is pointed out in Ref.~\citen{KBS}.

\bibitem{note:Burnett}
Some Burnett coefficients might diverge as in the Lorentz model \cite{EMDB1}.

\bibitem{OH}
M. Otsuki and H. Hayakawa, arXiv:0711.1421.

\end{thebibliography}
\end{document}